\documentclass[a4paper,papersize,11pt]{article}
\usepackage{graphics}
\usepackage[dvipdfmx]{graphicx}
\usepackage{amsmath}
\usepackage{amssymb}
\usepackage{amscd}
\usepackage[mathscr]{eucal}
\usepackage{bm}
\usepackage{simplewick}

\def\normOrd#1{\mathop{:}\nolimits\!#1\!\mathop{:}\nolimits}
\def\normOrdx#1{\mathop{:}\nolimits\!#1\!\mathop{:}\nolimits}

\begin{document}
\title{A New Generalized Wick Theorem
in Conformal Field Theory}
\author{Taichiro Takagi\\
\normalsize
\em Department of Applied Physics, National Defense Academy,
Kanagawa 239-8686, Japan}

\date{}
\maketitle
\thispagestyle{empty}
\begin{abstract}
A new generalized Wick theorem for interacting fields in 2D conformal field theory
is described.
We briefly discuss its relation to the Borcherds identity and its derivation by
an analytic method.
Examples of the calculations of the operator product expansions
by using the generalized Wick theorems including fermionic fields
are also presented.
\end{abstract}
\section{Introduction}\label{sec:1}

The well known generalized Wick theorem for interacting fields in two dimensional conformal field theory (CFT) is given as follows.
It was originally presented by Bais, Bouwknegt, Surridge and Schoutens \cite{BBSS88} in 1988,
and described more precisely in a standard textbook of conformal field theory \cite{DMS97}. 
\begin{equation}
\contraction{}{A}{(z)(}{BC}
A(z)(BC)(w) 
= \frac{1}{2 \pi \sqrt{-1}}
\oint_{C_w} \frac{d x}{x-w} \left\{\contraction{}{A}{(z)}{B}
A(z)B(x) C(w)+ B(x) \contraction{}{A}{(z)}{C}A(z) C(w) \right\}. \label{eq:july3_1}
\end{equation}
Here, $A(z)$ etc.~are operators for chiral conformal fields,
$(BC)(w) = \normOrd{B(w)C(w)}$ denotes the normally ordered product,
{\scriptsize $\contraction{}{A}{(z)}{B}A(z)B(x)$} denotes the contraction, i.~e.~the singular part of the operator product expansion (OPE), and
$C_w$ is a contour encircling the point $w$ with an infinitesimal radius.

 It is a natural question to ask whether an analogous expression exists
for the contraction {\scriptsize $\contraction{}{(AB)}{(z)(}{C}(AB)(z)C(w)$ }
in which a normally ordered operator is on the left and
a single operator is on the right.
As far as the author knows, no one had correctly answered this question yet.
So we tried to answer it and found the following expression \cite{TY16}.
\begin{equation}
\contraction{}{(AB)}{(z)(}{C}
(AB)(z)C(w) 
= \frac{1}{2 \pi \sqrt{-1}}
\oint_{C_w} \frac{d x}{z-x} \left\{
\normOrdx{A(x) \contraction{}{B}{(x)}{C}B(x)C(w)} + B(x) \contraction{}{A}{(x)}{C}A(x) C(w) \right\}. \label{eq:main}
\end{equation}
Here the normal ordering for the first term in the integrand
is for dropping the singular terms arising in the 
OPE of 
$A(x)$ and the coefficients of the contraction 
on its right hand.
We note that even in \eqref{eq:july3_1} one can replace the second term by
 {\scriptsize$\normOrd{B(x)\contraction{}{A}{(z)}{C}A(z) C(w)}$} 
which makes the two formulas apparently more analogous,
but usually do not do so
because the
singular terms are automatically dropped by the integral in \eqref{eq:july3_1}.
It is a rather remarkable fact that such a really simple formula 
\eqref{eq:main} has not been found until today
over a quarter of a century after \eqref{eq:july3_1}.
%

In section \ref{sec:2} we briefly describe a proof of the generalized 
Wick theorem \eqref{eq:main} based on
the Borcherds identity, and also sketch the idea of its derivation by an analytic method.
In section \ref{sec:3}  we present examples of the calculations of the operator product expansions
by using the generalized Wick theorems including fermionic fields.
For reasons of simplicity, I will omit almost all mathematically rigorous arguments
which are available in \cite{TY16} for interested readers.

%
\section{Wick Theorems as Special Cases of the Borcherds Identity}
\label{sec:2}
\subsection{Notations}
For a mathematical formulation of two-dimensional chiral quantum fields we adopt the one given by Matsuo and Nagatomo in 1999 \cite{MN97}.
We express the operator $A(y)$ as a formal power series
\begin{equation}
A(y) = \sum_{n \in \mathbb{Z}} A_n y^{-n-1}
=\left( \sum_{n \geq 0}+ \sum_{n < 0} \right) A_n y^{-n-1}
=A(y)_+ + A(y)_-
,\label{eq:july6_1}
\end{equation}
and divide it into its positive and negative parts in this way.
Here the coefficients $A_n$ are linear transformations
on some vector space $M$.

We define the normally ordered product of $A(y)$ and $B(z)$ as follows
\begin{equation}
\normOrd{A(y)B(z)} = A(y)_{-} B(z) + B(z) A(y)_{+}.
\label{eq:july3_3}
\end{equation}
Then it turns out to be regular at $y=z$ in a sense.

Now we want to introduce an important notion called the $n$th residue product.
Its definition is given by the following formula for non-negative $n$.
\begin{equation}
A(z)_{(n)}B(z) =
{\rm Res}_{y=0} [A(y), B(z)] (y-z)^n.
\label{eq:oct20_1}
\end{equation}
Then, we define the contraction as follows.
\begin{equation}
\contraction{}{A}{(y)}{B} A(y)B(z) =  \sum_{n=0}^{\infty} \frac{A(z)_{(n)}B(z)}{(y-z)^{n+1}}.
\label{eq:july3_2}
\end{equation}
And we have such an operator product expansion 
\begin{equation}\label{eq:july29_3}
A(y)B(z) = \contraction{}{A}{(y)}{B} A(y)B(z) + \normOrd{A(y)B(z)},
\end{equation}
in the region $|y| > |z|$.
By slightly modifying the expression in \eqref{eq:oct20_1}, we can define the residue product
also for negative $n$.
In particular, we frequently use the fact that the minus first residue product is
equal to the normally ordered product
\begin{equation*}
A(z)_{(-1)}B(z) =\normOrd{A(z)B(z)}.
\end{equation*}
We note that sometimes we write the definition simply as \eqref{eq:oct20_1} even for negative $n$ just
as an abbreviation. 

\subsection{Borcherds identity}\label{sec:2_5}
Now we want to introduce the Borcherds identity.
First we write the usual Jacobi identity in this way
\begin{equation}
[[A,B],C] = [A,[B,C]] - [B,[A,C]].
\end{equation}
From this Jacobi identity, Matsuo and Nagatomo
derived such an identity.
\begin{eqnarray}
&&\sum_{i=0}^\infty {p \choose i} (A_{(r+i)}B)_{(p+q-i)}C
=\sum_{i=0}^\infty (-1)^i  {r \choose i}
\left( A_{(p+r-i)}(B_{(q+i)} C) 
-(-1)^r
B_{(q+r-i)} (A_{(p+i)} C) \right).\nonumber\\
&&\label{eq:borcherds}
\end{eqnarray}
Here we omitted the arguments of the operators
that assumed to be $w$ in what follows.
The small brackets in the subscripts
mean that we are taking the residue products twice.
This is the so-called Borcherds identity
known as one of the axioms for a vertex algebra \cite{B86}.
Matsuo and Nagatomo proved that it is valid for any integer $p,q,r$, 
under the assumption of a property called the locality of quantum fields.

\subsection{Non-commutative Wick formula by Kac}
Now let us consider a specialization of this identity by setting 
$r=0$ and $q=-1$ for non-negative $p$. 
\begin{equation}\label{eq:mar31_1}
A_{(p)} (B_{(-1)}C) =
 (A_{(p)} B)_{(-1)} C
+B_{(-1)} ( A_{(p)} C  )
+\sum_{i=0}^{p-1}  {p \choose i }
 (A_{(i)} B)_{(p-i-1)} C .
\end{equation}
In his famous book \cite{Kac96}, Kac derived this identity by his own
method and called it the ``non-commutative" Wick formula.
As he pointed out, for free fields the last term vanishes and it
reduces to the usual Wick theorem for free fields.
He also pointed out its equivalence to the generalized Wick theorem \eqref{eq:july3_1}.
A proof of this equivalence is available in \cite{TY16}

\subsection{Algebraic proof of the new generalized Wick theorem}
Now we want to show that our new generalized Wick 
theorem \eqref{eq:main} is equivalent to
another specialization of the Borcherds identity. 
To begin with we prepare some formulas.
One easily proves
\begin{equation*}
\partial A(w)_{(n)} B(w) = -n A(w)_{(n-1)}B(w),\label{eq:july29_2}
\end{equation*}
for any integer $n$.
By changing $n$ by $-n$ and using this relation repeatedly we obtain 
\begin{align*}
n! A_{(-n-1)}B &= (n-1) ! \partial A_{(-n)}B \\
 &= (n-2) ! \partial^2 A_{(-n+1)}B \\
 &= \cdots \\
&= \partial^n A_{(-1)}B,
\end{align*}
hence
\begin{equation}\label{eq:sept8_1}
\normOrd{\partial^{(n)}A B} = A_{(-n-1)}B,
\end{equation}
for any
non-negative integer $n$.
Here $\partial^{(n)} = \partial^n/n!$.
Now the first term in the integral of our formula $\normOrd{A(x) \contraction{}{B}{(x)}{C}B(x)C(w)}$ can be written as follows.
\begin{equation}\label{eq:apr4_1}
\normOrdx{\,A(x) \contraction{}{B}{(x)}{C} B(x)C(w)}=
\sum_{i=0}^{\infty} \frac{\normOrd{ \,A(x) (B(w)_{(i)}C(w))}}{(x-w)^{i+1}}
= 
\sum_{i=0}^{\infty} \sum_{j=0}^{\infty}
\frac{A_{(-j-1)} (B_{(i)}C)}{(x-w)^{i-j+1}}.
\end{equation}
Here we replaced $A(x)$ by its Taylor expansion around $w$ and then used the above formula \eqref{eq:sept8_1}.
By multiplying $(z-x)^{-1}$ and integrating over $x$ around $w$,
only the singular terms survive, and then all the $x$'s are replaced by $z$'s.
Then by setting $i=q+j$ we obtain
\begin{equation}
\frac{1}{2 \pi \sqrt{-1}}
\oint_{C_w} \frac{d x}{z-x} \left\{
\normOrdx{A(x) \contraction{}{B}{(x)}{C}B(x)C(w)} \right\}
=
\sum_{i=0}^{\infty} \sum_{j=0}^{i}
\frac{A_{(-j-1)} (B_{(i)}C)}{(z-w)^{i-j+1}}
=
\sum_{q=0}^{\infty} \sum_{j=0}^{\infty}
\frac{A_{(-j-1)} (B_{(q+j)}C)}{(z-w)^{q+1}}.\label{eq:jan27_5}
\end{equation}

Now let us consider the second term of the integrand of our formula.
We divide it into the singular part and the regular part.
\begin{equation}
B(x) \contraction{}{A}{(x)}{C}A(x) C(w)  = \bcontraction{}{B}{(x) (A(x}{)}
\contraction{B(x) (}{A}{(x)}{C}
B(x) (A(x) C(w)) + \normOrdx{\,B(x) \contraction{}{A}{(x)}{C} A(x)C(w)}.
\end{equation}
By the same argument as above it turns out to be
\begin{equation}
\sum_{i=0}^{\infty} 
\sum_{j=0}^{\infty} 
\frac{B_{(j)} (A_{(i)}C)}{(z-w)^{i+j+2} }
+
\sum_{i=0}^{\infty} \sum_{j=0}^{i}
\frac{B_{(-j-1)} (A_{(i)}C)}{(z-w)^{i-j+1}},
\end{equation}
after the calculation of the integral.
We can put these terms together,
by deleting the second term and
by changing the lower bound 
of the summation over
$j$ in the first term by $-i-1$.
Then by setting $j=q-i-1$ we obtain
\begin{equation}
\frac{1}{2 \pi \sqrt{-1}}
\oint_{C_w} \frac{d x}{z-x} \left\{
B(x) \contraction{}{A}{(x)}{C}A(x) C(w) 
\right\}
=
\sum_{i=0}^{\infty} 
\sum_{j=-i-1}^{\infty} 
\frac{B_{(j)} (A_{(i)}C)}{(z-w)^{i+j+2} }
=
\sum_{q=0}^{\infty} 
\sum_{i=0}^{\infty} 
\frac{B_{(q-i-1)} (A_{(i)}C)}{(z-w)^{q+1} }.
\end{equation}
From these results we find that our new generalized Wick theorem \eqref{eq:main}
is equivalent to
\begin{equation}\label{eq:mar31_2}
(A_{(-1)}B)_{(q)}C=
\sum_{i=0}^{\infty}
\left( A_{(-i-1)} (B_{(q+i)}C)
+B_{(q-i-1)} (A_{(i)}C) \right).
\end{equation}
As you can see, this is nothing but a specialization of the Borcherds identity
with $p=0, r=-1$. 

\subsection{Analytic proof of the new generalized Wick theorem}
As we have just shown, once you have obtained the expression for our new
generalized Wick theorem, you can check its validity by simply calculating the integrals
and then by comparing the result with the Borcherds identity.
However, it is a natural question to ask how you can 
come up with such an expression.
In the remaining part of this section, we want to answer this question.
We are going to sketch the idea for deriving our new formula by using an analytic method.

By definition the nested residue product $(A_{(-1)}B)_{(p)}C$
is given by evaluating the residues twice for a nested commutator
of the operators.
We rewrite this expression by using contour integrals as
\begin{align*}
&(A_{(-1)} B)_{(p)}C)
={\rm Res}_{x=0} {\rm Res}_{y=0} 
 [[A(y), B(x)], C(w)]  (y-x)^{-1} (x-w)^{p}\\
&=\frac{1}{(2 \pi \sqrt{-1})^2}
\left\{ \oint_{C_{3}}\!\!\!d x \oint_{C_{4}}\!\!\!d y
-\oint_{C_{4}}\!\!\!d x \oint_{C_{3}}\!\!\!d y
-\oint_{C_{1}}\!\!\!d x \oint_{C_{2}}\!\!\!d y
+\oint_{C_{2}}\!\!\!d x \oint_{C_{1}}\!\!\!d y \right\}
 A(y)B(x)C(w)  \frac{(x-w)^{p}}{(y-x)}.
\end{align*}
Here, the four terms are arising from the expansion of the nested commutators.
The contours $C_{1,2,3,4}$ are those encircling the origin with increasing radii,
and the point $w$ is between $C_2$ and $C_3$.
\begin{figure}[hbtp]
\centering
\scalebox{1}[1]{
\includegraphics[height=6cm,clip]{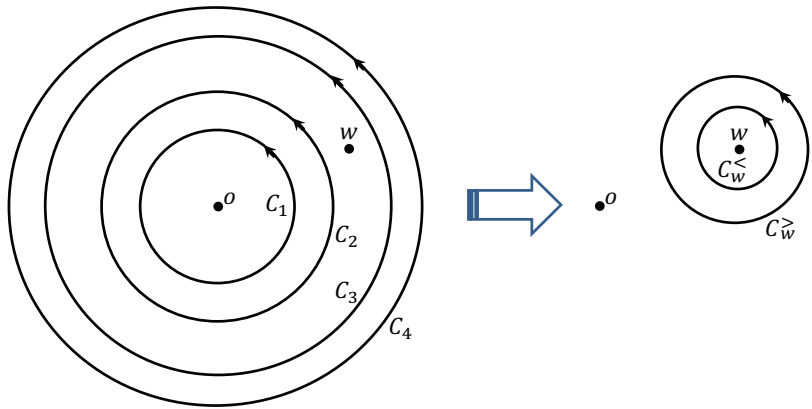}}
\caption{Deformation of the contours}
\label{f1}
\end{figure}
We let $C_3$ and $C_4$ shrink to $C_1$ and $C_2$ respectively,
so that they are going to cancel out each other. 
But of course by deforming them they pick up the singularity at the point $w$.
So we have the following expression.
\begin{equation*}
(A_{(-1)} B)_{(p)}C)
=\frac{1}{(2 \pi \sqrt{-1})^2}
\left\{ \oint_{C_{w}^{>}}\!\!\!d y \oint_{C_{w}^{<}}\!\!\!d x
-\oint_{C_{w}^{>}}\!\!\!d x \oint_{C_{w}^{<}}\!\!\!d y
\right\}
 A(y)B(x)C(w)  \frac{(x-w)^{p}}{(y-x)}.
\end{equation*}
Here, the contours $C_{w}^{>}$  and $C_{w}^{<}$ are given as in Fig.~\ref{f1}. 
By multiplying $(z-w)^{-p-1}$ and then taking a summation over $p$ from $0$ to
infinity, we obtain
\begin{equation*}
\contraction{}{(AB)}{(z)(}{C}
(AB)(z)C(w) 
=\frac{1}{(2 \pi \sqrt{-1})^2}
\left\{ \oint_{C_{w}^{>}}\!\!\!d y \oint_{C_{w}^{<}}\!\!\!d x
-\oint_{C_{w}^{>}}\!\!\!d x \oint_{C_{w}^{<}}\!\!\!d y
\right\}
 A(y)B(x)C(w)  \frac{1}{(y-x)(z-x)}.
\end{equation*}
Let us consider the second term first.
By taking the minus sign into account, its inner integral is given by this expression:
\begin{equation*}
\frac{1}{2 \pi \sqrt{-1}}
\oint_{C_{w}^{<}}\! \frac{d y}{x-y}
 A(y)B(x)C(w) =
B(x) \contraction{}{A}{(x)}{C}A(x) C(w).
\end{equation*}
Since this is an integral over the variable $y$ around the point $w$,
it picks up the singular part of the operator product expansion of $A(y)C(w)$
and then replaces $y$ by $x$, hence the above result.
After the outer integral over $x$ we
obtain the second term of our generalized Wick theorem.

Now let us consider the first term.
First we write this factor $(y-x)^{-1}(z-x)^{-1}$ as a difference like this:
\begin{equation*}
 \frac{1}{(y-x)(z-x)} =
 \frac{1}{z-y} \left( 
 \frac{1}{y-x}  -  \frac{1}{z-x} 
\right).
\end{equation*}
Then the inner integral over $x$ is expressed as
\begin{equation*}
\frac{1}{2 \pi \sqrt{-1}}
\left(
\oint_{C_{w}^{<}}\! \frac{d x}{y-x} -
\oint_{C_{w}^{<}}\! \frac{d x}{z-x}
\right)
 A(y)B(x)C(w) =
A(y) \contraction{}{B}{(y)}{C}B(y) C(w)-
A(y) \contraction{}{B}{(z)}{C}B(z) C(w).
\end{equation*}
Just like the previous calculation, we obtain the result of integrals
as above.
Now let us consider the outer integral over $y$.
\begin{equation*}
\frac{1}{2 \pi \sqrt{-1}}
\oint_{C_w} \frac{d y}{z-y} \left( A(y) \contraction{}{B}{(y)}{C}B(y) C(w)-
A(y) \contraction{}{B}{(z)}{C}B(z) C(w) \right)
=\frac{1}{2 \pi \sqrt{-1}}
\oint_{C_w} \frac{d y}{z-y} 
\normOrd{\,A(y) \contraction{}{B}{(y)}{C} B(y)C(w)} .
\end{equation*}
The regular part of the second term vanishes by itself because 
it has no singularity at $y=w$.
The singular parts of the first and the second terms cancel out each other
after integration.
Thus, there remains only the regular part of the first term as above.
By changing $y$ by $x$, we obtain the first term of our generalized Wick theorem
\eqref{eq:main}.

\subsection{Example 1: The free boson}
Let $J(z)$ be the current of a free bosonic field
satisfying
\begin{equation*}
J(z)J(w)  = \frac{1}{(z-w)^2} + \normOrd{J(z)J(w)},
\end{equation*}
and define the energy momentum tensor as
\begin{equation*}
T(z) = \frac12  \normOrd{J(z)J(z)} = \frac12 (JJ)(z).
\end{equation*}
Then by using \eqref{eq:july3_1} we have
\begin{align*}
\contraction{}{J}{(z)}{T}J(z)T(w) &=  \frac12 \contraction{}{J}{(z)(}{JJ}
J(z)(JJ)(w)  = \frac12 \cdot \frac{1}{2 \pi \sqrt{-1}}
\oint_{C_w} \frac{d x}{x-w} \left\{ \frac{1}{(z-x)^2} J(w) + J(x) \frac{1}{(z-w)^2} \right\}
\nonumber\\
&= \frac{J(w)}{(z-w)^2}.
\end{align*}
Now by \eqref{eq:main} we obtain
\begin{align*}
\contraction{}{T}{(z)}{T}T(z)T(w) &=  \frac12 
\contraction{(}{JJ}{)(z)}{T}
(JJ)(z)T(w) = \frac12 \cdot \frac{1}{2 \pi \sqrt{-1}}
\oint_{C_w} \frac{d x}{z-x} \left\{ \normOrd{J(x) \frac{J(w)}{(x-w)^2}} + J(x) \frac{J(w)}{(x-w)^2} \right\}\\
&= \frac12 \cdot \frac{1}{2 \pi \sqrt{-1}}
\oint_{C_w} \frac{d x}{z-x} \left\{ \frac{1}{(x-w)^4} + \frac{2 \normOrd{J(x)J(w)}}{(x-w)^2} \right\}\\
&= \frac12 \cdot \left\{ \frac{1}{(z-w)^4} + \frac{2 (JJ)(w)}{(z-w)^2} 
+ \frac{2 \normOrd{\partial J(w)J(w)}}{z-w}
\right\}\\
&=  \frac{1/2}{(z-w)^4} + \frac{2 T(w)}{(z-w)^2} 
+ \frac{\partial T(w)}{z-w}.
\end{align*}
This is the OPE for the CFT with the central charge $c=1$.
Here we used the relation $\normOrd{\partial J(w)J(w)} = \normOrd{J(w)\partial J(w)}$
that is due to the following skew symmetry \cite{MN97}
\begin{equation}\label{eq:skewsym}
B(z)_{(m)}A(z) = \sum_{i=0}^\infty (-1)^{m+i+1} \partial^{(i)}
(A(z)_{(m+i)}B(z) ).
\end{equation}

\section{Application to Operator Product Expansions
for Fermionic Fields}\label{sec:3}

\subsection{Extension of the generalized Wick theorems to including fermionic fields}
For any field $A$ let $p(A)$ be its parity, i.~e.~$p(A)=0$ (resp.~$p(A)=1$) if
$A$ is a bosonic (resp.~fermionic) field \cite{Kac96}.
Then the extension of the generalized Wick theorems to including fermionic fields
should be given by
\begin{align}
\contraction{}{A}{(z)(}{BC}
A(z)(BC)(w) 
&= \frac{1}{2 \pi \sqrt{-1}}
\oint_{C_w} \frac{d x}{x-w} \left\{\contraction{}{A}{(z)}{B}
A(z)B(x) C(w)+ (-1)^{p(A)p(B)}B(x) \contraction{}{A}{(z)}{C}A(z) C(w) \right\}, \label{eq:gwt1}\\
\contraction{}{(AB)}{(z)(}{C}
(AB)(z)C(w) 
&= \frac{1}{2 \pi \sqrt{-1}}
\oint_{C_w} \frac{d x}{z-x} \left\{
\normOrdx{A(x) \contraction{}{B}{(x)}{C}B(x)C(w)} + (-1)^{p(A)p(B)}B(x) \contraction{}{A}{(x)}{C}A(x) C(w) \right\}. \label{eq:gwt2}
\end{align}
In what follows we present two simplest non-trivial examples to
illustrate their validity. 

\subsection{Example 2: The free fermion}
Let $\psi (z)$ be the free fermion satisfying
\begin{equation}
\psi (z) \psi (w) = \frac{1}{z-w} + \normOrd{\psi (z) \psi (w)}.
\end{equation}
Define the energy momentum tensor $T(z)$ as
\begin{equation}
T(z) = -\frac12 \normOrd{\psi (z) \partial \psi (z)}.
\end{equation}
By \eqref{eq:gwt1} we have
\begin{align}
\contraction{}{\psi}{(z)}{T}
\psi (z)T(w) 
&= -\frac12 \frac{1}{2 \pi \sqrt{-1}}
\oint_{C_w} \frac{d x}{x-w} \left\{\contraction{}{\psi}{(z)}{\psi}
\psi(z)\psi(x) \partial \psi(w)-\psi(x) \contraction{}{\psi}{(z) \partial}{\psi}\psi(z) \partial \psi(w) \right\} \nonumber\\
&= -\frac12 \frac{1}{2 \pi \sqrt{-1}}
\oint_{C_w} \frac{d x}{x-w} \left\{\frac{\partial \psi(w)}{z-x}-\frac{\psi(x)}{(z-w)^2}  \right\} \nonumber\\
&= \frac{\frac12 \psi(w)}{(z-w)^2}-\frac{\frac12 \partial \psi(w)}{z-w},
\end{align}
and hence
\begin{equation}
\partial \contraction{}{\psi}{(z)}{T}
\psi (z)T(w) =- \frac{\psi(w)}{(z-w)^3}+\frac{\frac12 \partial \psi(w)}{(z-w)^2}.
\end{equation}
Now by \eqref{eq:gwt2} we have
\begin{align}
\contraction{}{T}{(z)(}{T}
T(z)T(w) 
&= -\frac12 \frac{1}{2 \pi \sqrt{-1}}
\oint_{C_w} \frac{d x}{z-x} \left\{
\normOrdx{\psi(x) \partial \contraction{}{\psi}{(x)}{T}\psi(x)T(w)} -\partial \psi(x) \contraction{}{\psi}{(x)}{T}\psi(x) T(w) \right\}. 
\end{align}
The two terms in the integrand can be written as
\begin{equation}
\normOrdx{\psi(x) \partial \contraction{}{\psi}{(x)}{T}\psi(x)T(w)}=
\frac{\normOrd{\psi (w) \psi (x)}}{(x-w)^3}
+\frac{\frac12 \normOrd{\psi (x) \partial \psi (w)}}{(x-w)^2},
\end{equation}
and
\begin{align}
 -\partial \psi(x) \contraction{}{\psi}{(x)}{T}\psi(x) T(w) 
&= \frac{-\frac12}{(x-w)^2} \partial \psi (x) \psi (w)
+ \frac{\frac12}{x-w} \partial \psi (x) \partial \psi (w)
\nonumber\\
&= \frac{-\frac12}{(x-w)^4}+
\frac{\frac12 \normOrd{\psi (w) \partial \psi (x)}}{(x-w)^2}+
\frac{\frac12 \normOrd{\partial \psi (x) \partial \psi (w)}}{x-w}.
\end{align}
By summing them up and expanding the numerators around $x-w$ we obtain
\begin{align}
&\normOrdx{\psi(x) \partial \contraction{}{\psi}{(x)}{T}\psi(x)T(w)} -\partial \psi(x) \contraction{}{\psi}{(x)}{T}\psi(x) T(w)\nonumber\\
&=\frac{-\frac12}{(x-w)^4}+
\frac{2 \normOrd{\psi (w) \partial \psi (w)}}{(x-w)^2}+
\frac{\normOrd{\psi (w) \partial^2 \psi (w)}}{x-w}+
\cdots ,
\end{align}
where ``$\cdots$" denote the terms that are regular at $x=w$.
Therefore
\begin{align}
\contraction{}{T}{(z)(}{T}
T(z)T(w) 
&=-\frac12
\left\{
\frac{-\frac12}{(z-w)^4}+
\frac{2 \normOrd{\psi (w) \partial \psi (w)}}{(z-w)^2}+
\frac{\normOrd{\psi (w) \partial^2 \psi (w)}}{z-w}
\right\}
\nonumber\\
&=\frac{\frac14}{(z-w)^4}+
\frac{2 T(w)}{(z-w)^2}+
\frac{\partial T(w)}{z-w}.
\end{align}
This is the OPE for the CFT with the central charge $c=\frac12$.

One can also use \eqref{eq:gwt2} to calculate
\begin{align}
\contraction{}{T}{(z)}{\psi}
T(z)\psi(w) 
&= -\frac12 \frac{1}{2 \pi \sqrt{-1}}
\oint_{C_w} \frac{d x}{z-x} \left\{
\normOrdx{\psi(x) \partial \contraction{}{\psi}{(x)}{\psi}\psi(x)\psi(w)} -\partial \psi(x) \contraction{}{\psi}{(x)}{\psi}\psi(x) \psi(w) \right\}\nonumber\\
&= \frac{\frac12 \psi(w)}{(z-w)^2}+\frac{\partial \psi(w)}{z-w},
\end{align}
implying that $\psi (w)$ is a primary field with conformal dimension $1/2$.
This is because the terms in the integrand can be written as
\begin{align}
\normOrdx{\psi(x) \partial \contraction{}{\psi}{(x)}{\psi}\psi(x)\psi(w)} -\partial \psi(x) \contraction{}{\psi}{(x)}{\psi}\psi(x) \psi(w)
&=
-\frac{\psi(x)}{(x-w)^2}-\frac{\partial \psi (x)}{x-w} \nonumber\\
&=
-\frac{\psi(w)}{(x-w)^2}-\frac{2 \partial \psi (w)}{x-w}+
\cdots .
\end{align}

\subsection{Example 3: The ghost system}
In this section we describe applications of the generalized Wick theorems
\eqref{eq:gwt1},  \eqref{eq:gwt2} to the so-called $(b,c)$ ghost system
\cite{BBS07, BP09, DMS97, Y06}.
Let $b,c$ be a pair of fermions satisfying
\begin{equation}
\contraction{}{b}{(z)}{b}b(z)b(w)=0,\quad
\contraction{}{c}{(z)}{c}c(z)c(w)=0,\quad
\contraction{}{b}{(z)}{c}b(z)c(w)=\frac{1}{z-w}.
\end{equation}
We define
\begin{equation}
J(z) = \normOrd{b(z)c(z)},\quad
A(z) = \normOrd{b'(z)c(z)},\quad
B(z) = - \normOrd{b(z) c'(z)}.
\end{equation}
By \eqref{eq:gwt1} we have
\begin{equation}
\contraction{}{b}{(z)}{J}b(z)J(w) = -\frac{b(w)}{z-w},\quad
\contraction{}{\vphantom{b}c}{(z)}{J}\vphantom{b}c(z)J(w) = \frac{c(w)}{z-w},
\label{eq:oct12_1}
\end{equation}
and also
\begin{equation}
\contraction{}{b}{(z)}{A}b(z)A(w) = -\frac{b'(w)}{z-w},\quad
\contraction{}{b}{(z)}{B}b(z)B(w) = \frac{b(w)}{(z-w)^2},\quad
\contraction{}{\vphantom{b}c}{(z)}{A}\vphantom{b}c(z)A(w) = \frac{c(w)}{(z-w)^2},\quad
\contraction{}{\vphantom{b}c}{(z)}{B}\vphantom{b}c(z)B(w) = -\frac{c'(w)}{z-w}.
\label{eq:oct12_2}
\end{equation}
Then by \eqref{eq:gwt2} and \eqref{eq:oct12_1} we have
\begin{align}
\contraction{}{J}{(z)(}{J}
J(z)J(w) 
&=\frac{1}{2 \pi \sqrt{-1}}
\oint_{C_w} \frac{d x}{z-x} \left\{
\normOrdx{b(x) \contraction{}{\vphantom{b}c}{(x)}{J}\vphantom{b}c(x)J(w)} -c(x) \contraction{}{b}{(x)}{J}b(x) J(w) \right\} \nonumber\\
&=\frac{1}{2 \pi \sqrt{-1}}
\oint_{C_w} \frac{d x}{z-x} \left\{
\frac{\normOrd{b(x)c(w)}}{x-w} +\frac{c(x)b(w) }{x-w } \right\}\nonumber\\
&=\frac{1}{2 \pi \sqrt{-1}}
\oint_{C_w} \frac{d x}{z-x} \left\{\frac{1}{(x-w)^2}+
\frac{\normOrd{b(x)c(w)} + \normOrd{c(x)b(w)}}{x-w}  \right\}\nonumber\\
&=\frac{1}{2 \pi \sqrt{-1}}
\oint_{C_w} \frac{d x}{z-x} \left\{\frac{1}{(x-w)^2}+
\frac{\normOrd{b(w)c(w)} + \normOrd{c(w)b(w)}}{x-w} + \cdots \right\}\nonumber\\
&=\frac{1}{(z-w)^2}.
\end{align}
Thus the field $J(z)$ is a free bosonic current.

Also by \eqref{eq:gwt2} we have
\begin{align}
\contraction{}{A}{(z)(}{b}
A(z)b(w) 
&=\frac{1}{2 \pi \sqrt{-1}}
\oint_{C_w} \frac{d x}{z-x} \left\{
\normOrdx{b'(x) \contraction{}{\vphantom{b}c}{(x)}{b}\vphantom{b}c(x)b(w)} -c(x) \contraction{}{b'}{(x)}{b}b'(x) b(w) \right\} \nonumber\\
&=\frac{1}{2 \pi \sqrt{-1}}
\oint_{C_w} \frac{d x}{z-x} \left\{
\frac{b'(x)}{x-w}\right\}=\frac{b'(w)}{z-w},\\
\contraction{}{B}{(z)(}{b}
B(z)b(w) 
&=\frac{-1}{2 \pi \sqrt{-1}}
\oint_{C_w} \frac{d x}{z-x} \left\{
\normOrdx{b(x) \contraction{}{c'}{(x)}{b}c'(x)b(w)} -c'(x) \contraction{}{b}{(x)}{b}b(x) b(w) \right\} \nonumber\\
&=\frac{1}{2 \pi \sqrt{-1}}
\oint_{C_w} \frac{d x}{z-x} \left\{
\frac{b(x)}{(x-w)^2}\right\}=\frac{b(w)}{(z-w)^2} +\frac{b'(w)}{z-w},\\
\contraction{}{A}{(z)(}{c}
A(z)c(w) 
&=\frac{1}{2 \pi \sqrt{-1}}
\oint_{C_w} \frac{d x}{z-x} \left\{
\normOrdx{b'(x) \contraction{}{c}{(x)}{c}c(x)c(w)} -c(x) \contraction{}{b'}{(x)}{c}b'(x) c(w) \right\} \nonumber\\
&=\frac{1}{2 \pi \sqrt{-1}}
\oint_{C_w} \frac{d x}{z-x} \left\{
\frac{c(x)}{(x-w)^2}\right\}=\frac{c(w)}{(z-w)^2} +\frac{c'(w)}{z-w},\\
\contraction{}{B}{(z)(}{c}
B(z)c(w) 
&=\frac{-1}{2 \pi \sqrt{-1}}
\oint_{C_w} \frac{d x}{z-x} \left\{
\normOrdx{b(x) \contraction{}{c'}{(x)}{c}c'(x)c(w)} -c'(x) \contraction{}{b}{(x)}{c}b(x) c(w) \right\} \nonumber\\
&=\frac{1}{2 \pi \sqrt{-1}}
\oint_{C_w} \frac{d x}{z-x} \left\{
\frac{c'(x)}{x-w}\right\}=\frac{c'(w)}{z-w}.
\end{align}
In the same way we have
\begin{align}
\contraction{}{A}{(z)(}{J}
A(z)J(w) 
&=\frac{1}{2 \pi \sqrt{-1}}
\oint_{C_w} \frac{d x}{z-x} \left\{
\normOrdx{b'(x) \contraction{}{\vphantom{b}c}{(x)}{J}\vphantom{b}c(x)J(w)} -c(x) \contraction{}{b'}{(x)}{J}b'(x) J(w) \right\} \nonumber\\
&=\frac{1}{2 \pi \sqrt{-1}}
\oint_{C_w} \frac{d x}{z-x} \left\{
\frac{\normOrd{b'(x)c(w)}}{x-w} -\frac{c(x)b(w) }{(x-w)^2 } \right\}\nonumber\\
&=\frac{1}{2 \pi \sqrt{-1}}
\oint_{C_w} \frac{d x}{z-x} \left\{
\frac{\normOrd{b'(x)c(w)}}{x-w} -\frac{1}{(x-w)^3}
+\frac{\normOrd{b(w)c(x)} }{(x-w)^2 } \right\}\nonumber\\
&=\frac{1}{2 \pi \sqrt{-1}}
\oint_{C_w} \frac{d x}{z-x} \left\{
-\frac{1}{(x-w)^3}
+\frac{\normOrd{b(w)c(w)} }{(x-w)^2 } 
+\frac{\normOrd{b'(w)c(w)}+\normOrd{b(w)c'(w)}}{x-w}+\cdots
\right\}\nonumber\\
&=\frac{-1}{(z-w)^3} + \frac{J(w)}{(z-w)^2} + \frac{J'(w)}{z-w},
\end{align}
and
\begin{align}
\contraction{}{B}{(z)(}{J}
B(z)J(w) 
&=\frac{1}{2 \pi \sqrt{-1}}
\oint_{C_w} \frac{d x}{z-x} \left\{
-\normOrdx{b(x) \contraction{}{c'}{(x)}{J}c'(x)J(w)} +c'(x) \contraction{}{b}{(x)}{J}b(x) J(w) \right\} \nonumber\\
&=\frac{1}{2 \pi \sqrt{-1}}
\oint_{C_w} \frac{d x}{z-x} \left\{
\frac{\normOrd{b(x)c(w)} }{(x-w)^2 }
-\frac{c'(x)b(w)}{x-w} \right\}\nonumber\\
&=\frac{1}{2 \pi \sqrt{-1}}
\oint_{C_w} \frac{d x}{z-x} \left\{
\frac{\normOrd{b(x)c(w)} }{(x-w)^2 }
+\frac{1}{(x-w)^3}-\frac{\normOrd{c'(x)b(w)}}{x-w} \right\}\nonumber\\
&=\frac{1}{2 \pi \sqrt{-1}}
\oint_{C_w} \frac{d x}{z-x} \left\{
\frac{1}{(x-w)^3}
+\frac{\normOrd{b(w)c(w)} }{(x-w)^2 } 
+\frac{\normOrd{b'(w)c(w)}+\normOrd{b(w)c'(w)}}{x-w}+\cdots
\right\}\nonumber\\
&=\frac{1}{(z-w)^3} + \frac{J(w)}{(z-w)^2} + \frac{J'(w)}{z-w}.
\end{align}
For any $\lambda \in \mathbb{C}$ let
\begin{align}
T(z) &= (1-\lambda)A(z) + \lambda B(z) 
=(1-\lambda) \normOrd{b'(z)c(z)} -\lambda \normOrd{b(z) c'(z)}.
\label{eq:oct12_3}
\end{align}
Then we have 
\begin{align}
\contraction{}{T}{(z)(}{b}T(z) b(w) &= \frac{\lambda}{(z-w)^2} b(w) + \frac{1}{z-w} b'(w),
\nonumber\\
\contraction{}{T}{(z)(}{c}T(z) c(w) &= \frac{(1- \lambda)}{(z-w)^2} c(w) + \frac{1}{z-w} c'(w),
\end{align}
and also
\begin{equation}
\contraction{}{T}{(z)(}{J}T(z) J(w) 
= \frac{2\lambda -1}{(z-w)^3} + \frac{1}{(z-w)^2} J(w) + \frac{1}{z-w} J'(w).
\end{equation}
Therefore with regard to the above energy momentum tensor $T(z)$
the fields $b,c$ behave as the primary fields with conformal dimension $\lambda, 1-\lambda$ respectively.
The current $J$ has conformal dimension $1$, but is not a primary field
unless $\lambda = 1/2$ \cite{Y06}.

Now by \eqref{eq:gwt2} and \eqref{eq:oct12_2} we have
\begin{align}
\contraction{}{A}{(z)(}{A}
A(z)A(w) 
&=\frac{1}{2 \pi \sqrt{-1}}
\oint_{C_w} \frac{d x}{z-x} \left\{
\normOrdx{b'(x) \contraction{}{\vphantom{b}c}{(x)}{A}\vphantom{b}c(x)A(w)} -c(x) \contraction{}{b'}{(x)}{A}b'(x) A(w) \right\} \nonumber\\
&=\frac{1}{2 \pi \sqrt{-1}}
\oint_{C_w} \frac{d x}{z-x} \left\{
\frac{\normOrd{b'(x)c(w)}}{(x-w)^2} -\frac{c(x)b'(w) }{(x-w)^2 } \right\}\nonumber\\
&=\frac{1}{2 \pi \sqrt{-1}}
\oint_{C_w} \frac{d x}{z-x} \left\{- \frac{1}{(x-w)^4}+
\frac{\normOrd{b'(x)c(w)} + \normOrd{b'(w)c(x)}}{(x-w)^2}  \right\}\nonumber\\
&=\frac{1}{2 \pi \sqrt{-1}}
\oint_{C_w} \frac{d x}{z-x} \left\{- \frac{1}{(x-w)^4}+
\frac{2 \normOrd{b'(w)c(w)}}{(x-w)^2} +
\frac{\normOrd{b''(w)c(w)} + \normOrd{b'(w)c'(w)}}{x-w}
+\cdots  \right\}\nonumber\\
&=-\frac{1}{(z-w)^4} + \frac{2 A(w)}{(z-w)^2}+\frac{A'(w)}{z-w},
\end{align}
\begin{align}
\contraction{}{A}{(z)(}{B}
A(z)B(w) 
&=\frac{1}{2 \pi \sqrt{-1}}
\oint_{C_w} \frac{d x}{z-x} \left\{
\normOrdx{b'(x) \contraction{}{\vphantom{b}c}{(x)}{B}\vphantom{b}c(x)B(w)} -c(x) \contraction{}{b'}{(x)}{B}b'(x) B(w) \right\} \nonumber\\
&=\frac{1}{2 \pi \sqrt{-1}}
\oint_{C_w} \frac{d x}{z-x} \left\{
-\frac{\normOrd{b'(x)c'(w)}}{x-w} +\frac{2 c(x)b(w) }{(x-w)^3 } \right\}\nonumber\\
&=\frac{1}{2 \pi \sqrt{-1}}
\oint_{C_w} \frac{d x}{z-x} \left\{\frac{2}{(x-w)^4}-
\frac{2 \normOrd{b(w)c(x)} }{(x-w)^3 }
-\frac{\normOrd{b'(x)c'(w)}}{x-w}  \right\}
\nonumber\\
&=\frac{1}{2 \pi \sqrt{-1}}
\oint_{C_w} \frac{d x}{z-x} \left\{\frac{2}{(x-w)^4}-
\frac{2 \normOrd{b(w)c(w)} }{(x-w)^3 }-
\frac{2 \normOrd{b(w)c'(w)} }{(x-w)^2 } \right. \nonumber\\
& \left. \qquad \qquad \qquad \qquad \qquad \qquad  -\frac{\normOrd{b'(w)c'(w)}+\normOrd{b(w)c''(w)}}{x-w} +\cdots \right\}
\nonumber\\
&=\frac{2}{(z-w)^4} -\frac{2J(w)}{(z-w)^3} 
+ \frac{2 B(w)}{(z-w)^2}+\frac{B'(w)}{z-w},
\end{align}
\begin{align}
\contraction{}{B}{(z)(}{A}
B(z)A(w) 
&=\frac{1}{2 \pi \sqrt{-1}}
\oint_{C_w} \frac{d x}{z-x} \left\{
-\normOrdx{b(x) \contraction{}{\vphantom{b}c'}{(x)}{A}\vphantom{b}c'(x)A(w)} +c'(x) \contraction{}{b}{(x)}{A}b(x) A(w) \right\} \nonumber\\
&=\frac{1}{2 \pi \sqrt{-1}}
\oint_{C_w} \frac{d x}{z-x} \left\{
\frac{2 \normOrd{b(x)c(w)} }{(x-w)^3}
-\frac{c'(x)b'(w)}{x-w}  \right\}\nonumber\\
&=\frac{1}{2 \pi \sqrt{-1}}
\oint_{C_w} \frac{d x}{z-x} \left\{\frac{2}{(x-w)^4}+
\frac{2 \normOrd{b(x)c(w)} }{(x-w)^3 }
+\frac{\normOrd{b'(w)c'(x)}}{x-w}  \right\}
\nonumber\\
&=\frac{1}{2 \pi \sqrt{-1}}
\oint_{C_w} \frac{d x}{z-x} \left\{\frac{2}{(x-w)^4}+
\frac{2 \normOrd{b(w)c(w)} }{(x-w)^3 }+
\frac{2 \normOrd{b'(w)c(w)} }{(x-w)^2 } \right. \nonumber\\
& \left. \qquad \qquad \qquad \qquad \qquad \qquad  +\frac{\normOrd{b''(w)c(w)}+\normOrd{b'(w)c'(w)}}{x-w} +\cdots \right\}
\nonumber\\
&=\frac{2}{(z-w)^4} +\frac{2J(w)}{(z-w)^3} 
+ \frac{2 A(w)}{(z-w)^2}+\frac{A'(w)}{z-w},
\end{align}
and
\begin{align}
\contraction{}{B}{(z)(}{B}
B(z)B(w) 
&=\frac{1}{2 \pi \sqrt{-1}}
\oint_{C_w} \frac{d x}{z-x} \left\{
-\normOrdx{b(x) \contraction{}{\vphantom{b}c'}{(x)}{B}\vphantom{b}c'(x)B(w)} +c'(x) \contraction{}{b}{(x)}{B}b(x) B(w) \right\} \nonumber\\
&=\frac{1}{2 \pi \sqrt{-1}}
\oint_{C_w} \frac{d x}{z-x} \left\{
-\frac{\normOrd{b(x)c'(w)}}{(x-w)^2} +\frac{c'(x)b(w) }{(x-w)^2 } \right\}\nonumber\\
&=\frac{1}{2 \pi \sqrt{-1}}
\oint_{C_w} \frac{d x}{z-x} \left\{- \frac{1}{(x-w)^4}-
\frac{\normOrd{b(x)c'(w)} + \normOrd{b(w)c'(x)}}{(x-w)^2}  \right\}\nonumber\\
&=\frac{1}{2 \pi \sqrt{-1}}
\oint_{C_w} \frac{d x}{z-x} \left\{- \frac{1}{(x-w)^4}-
\frac{2 \normOrd{b(w)c'(w)}}{(x-w)^2} -
\frac{\normOrd{b'(w)c'(w)} + \normOrd{b(w)c''(w)}}{x-w}
+\cdots  \right\}\nonumber\\
&=-\frac{1}{(z-w)^4} + \frac{2 B(w)}{(z-w)^2}+\frac{B'(w)}{z-w}.
\end{align}
Therefore the energy-momentum tensor \eqref{eq:oct12_3} satisfies
\begin{align}
\contraction{}{T}{(z)(}{T}
T(z)T(w) 
=\frac{\frac{c}{2}}{(z-w)^4}+
\frac{2 T(w)}{(z-w)^2}+
\frac{\partial T(w)}{z-w},
\end{align}
with the central charge $c=-2(6 \lambda^2 - 6 \lambda +1)$ \cite{Y06}.

\section{Concluding Remarks}\label{sec:4}
We described several features of our new generalized Wick theorem \eqref{eq:main} for interacting fields in 2D conformal field theory and presented its extension \eqref{eq:gwt2} for
including fermions.
Though the results in Examples 1, 2, 3 can also be derived by only using
the Wick theorem for free (non-interacting) fields, 
we presented them to
illustrate the validity of our generalized Wick theorems
which may be useful for readers who want to use our formulas extensively.
Since their expressions are so simple,
it is evident that there will be much more applications of our formulas
to various non-trivial problems in 2D conformal field theory.

\vspace{5mm}
\noindent
{\it Acknowledgement}.
This work was supported by 
JSPS KAKENHI Grant Number  JP25400122.

%

\end{document}